\documentclass[pra,twocolumn,graphicx,amssymb,floatfix]{revtex4}
\usepackage{graphicx}
\begin{document}

\title{ Comment on ``Experimental demonstration of direct path state characterization by strongly measuring weak values in a matter-wave interferometer'' }

\begin{abstract}
   It is argued that the strong coupling version of recent experiment [Denkmayr et al., PRL   118, 010402 (2017)] while correctly estimating the pre-selected states of the neutrons does not perform strong measurements of weak values as claimed.
        \end{abstract}
\maketitle

 Denkmayr {\it et al.} \cite{Denk} reported an experiment in which a tomographic task of  ``direct path state characterization''   in the neutron interferometer has been performed using weak and strong coupling to neutron's spin.  I correct  misleading statements in the title, abstract and conclusions regarding  strong measurements of weak values.

According to the title, direct path state characterization has been achieved by ``strongly measuring weak values". In the abstract: ``weak measurements are not a necessary condition to determine the weak value''. In the conclusions: ``we have presented a weak value determination scheme via arbitrary interaction strengths. We have applied it to experimentally determine weak values using both weak and strong interactions.'' I argue that in the strong regime, the experiment  does not measure weak values of the observed quantum system.

 The weak value of a  variable $A$ is a property of a quantum system at a particular time \cite{AAV}. It is specified by the forward and backward evolving quantum states at this time and it has a well defined operational meaning: any weak enough coupling to $A$   is an effective coupling to the weak value $A_w$. The pointer of a weakened von Neumann measurement is shifted in proportion to ${\rm Re} A_w$ while the  shift of the conjugate pointer variable is proportional to ${\rm Im} A_w$.

 Lundeen {\it et al.} \cite{Lund} pointed out that, given a particular post-selection,  the weak values of local projections are proportional to the local values of  the wave function and thus,  measurements of these weak values provide a  ``direct  measurement of the quantum wavefunction''. Vallone and Dequal \cite{Vall}  showed  that a modification  of this procedure, in which the weak coupling is replaced by a strong coupling, provides a more efficient method for ``direct  measurement of the quantum wavefunction'', although one might argue that it is less ``direct'', because instead of simple proportionality, we need  calculations to obtain the local amplitude from  a set of pointer  readings.

 Denkmayr {\it et al.} implemented these proposals in neutron interferometry, successfully accomplishing both strong and weak coupling versions of  the  ``path state characterization''. However, the strong coupling version of their experiment is not  a strong measurement of weak values as they claim.

 In the experiment, polarized neutrons, $|{\uparrow}_x\rangle$, are  prepared in  the path state $|P_i\rangle = a |{\rm I}\rangle+b |{\rm II}\rangle$ and  post-selected in   $|P_f\rangle = \frac{1}{\sqrt 2} (|{\rm I}\rangle+|{\rm II}\rangle)$. The task is to determine  $|P_i\rangle$.
  Weak values of the projections on the paths are $({\rm {\bf P}_I})_w=\frac{a}{a+b}$ and $({\rm {\bf P}_{II}})_w=\frac{b}{a+b}$.
   Proportionality of weak values to the complex amplitudes in the paths makes weak measurements of the projections ``direct'' measurements of the path state.

In a  more direct version of their experiment,  the polarization is rotated in one of the arms of the interferometer: $|{\uparrow_x}\rangle \rightarrow \cos \alpha |{\uparrow}_x\rangle - i\sin \alpha |{\downarrow}_x\rangle$. The spin tomography of the  output beam provides the information about  $|P_i\rangle$.
If we choose a small coupling, say in  path I, the angle of rotation in the $xy$ plane is  $2\alpha{\rm Re}({\rm {\bf P}_{\rm I}})_w$ and in the $xz$ plane   $2\alpha{\rm Im}({\rm {\bf P}_{\rm I}})_w$. After repeating the procedure  in path II,  $({\rm {\bf P}_I})_w$ and $({\rm {\bf P}_{II}})_w$  yield the pre-selected path state $|P_i\rangle$. In fact, since  $({\rm {\bf P}_I})_w +({\rm {\bf P}_{II}})_w =  ({\rm {\bf P}_I} +{\rm {\bf P}_{II}})_w=1$, we can   calculate $({\rm {\bf P}_{\rm II}})_w $  and the second procedure is not needed.

The first procedure with strong coupling (large $\alpha$) provides $|P_i\rangle$ even more efficiently \cite{Vall}. But does it measure the weak values of the projection, as the authors \cite{Denk} claim?

When the experiment runs with large  $\alpha$,  the two-state vector description of the neutrons inside the interferometer is different. The weak values of projections remain constant in  time, but their values are not the same as in the  run with vanishing polarization rotation.
 At the time before the post-selection, the state of the neutron is:
 $
|\Psi'\rangle = a |{\rm I}\rangle |{\uparrow}_x\rangle + b |{\rm II}\rangle ( \cos \alpha |{\uparrow}_x\rangle- i\sin \alpha |{\downarrow}_x\rangle).
$
Then, the neutron is partially post-selected onto path state
 $|P_f\rangle$. In such a case, the weak value is given by (13.23) of \cite{AV2008}
 \begin{equation}\label{psi}
  ({\rm {\bf P}_{I}})_w=\frac{\langle \Psi'|{\rm {\bf P}_{P_f}}{\rm {\bf P}_{I}}|\Psi'\rangle}{\langle  \Psi'| {\rm {\bf P}_{P_f}}|\Psi'\rangle}=\frac{a(b^\ast \cos \alpha+a^\ast)}{b(a^\ast \cos \alpha+b^\ast)+a(b^\ast \cos \alpha+a^\ast)}.
   \end{equation}
   The ratio of weak values of the projections,
  $ \frac{({\rm {\bf P}_{I}})_w}{({\rm {\bf P}_{II}})_w}=
   \frac{a(b^\ast \cos \alpha+a^\star)}{b(a^\ast \cos \alpha+b^\ast)},$
    yields the ratio of complex amplitudes only for vanishing interaction, $\alpha \rightarrow 0$.
     Direct path state characterization has not been done by strongly measuring weak values in this experiment because weak values cannot be measured strongly.

This work has been supported in part by the Israel Science Foundation Grant No. 1311/14,
the German-Israeli Foundation for Scientific Research and Development Grant No. I-1275-303.14.

L. Vaidman\\
 Raymond and Beverly Sackler School of Physics and Astronomy\\
 Tel-Aviv University, Tel-Aviv 69978, Israel

\end{document}